\documentclass[12pt,a4paper]{article}

\usepackage{amssymb}
\usepackage{amsthm}
\usepackage{graphicx}

\usepackage{caption}

\usepackage{color}
\usepackage{ulem}

\begin{document}

\title{Unbinding and unfolding of adhesion protein complexes through stretching:
interplay between shear and tensile mechanical clamps \\ 
\large{Protein complexes in mechanical clamps} }

\author{Bartosz R\'{o}\.{z}ycki$^{1,3}$, {\L}ukasz Mioduszewski$^2$ and Marek Cieplak$^1$ \\
$^1$Institute of Physics, Polish Academy of Sciences, \\ Al. Lotnik{\'o}w 32/46, 02-668 Warsaw, Poland \\
$^2$Faculty of Physics, University of Warsaw, \\ Ul. Ho\.za 69, 00-681 Warsaw, Poland \\
$^3$Correspondence to: Bartosz R\'{o}\.{z}ycki \\
Institute of Physics, Polish Academy of Sciences \\
Al. Lotnik{\'o}w 32/46, Warsaw 02-668, Poland \\
E-mail: rozycki@ifpan.edu.pl }

%\date{\today}

\date{August 21, 2014}

\maketitle

\clearpage

\begin{abstract}
\noindent 
Using coarse-grained molecular dynamics simulations, 
we analyze mechanically induced dissociation and unfolding of 
the protein complex CD48-2B4. This heterodimer is an indispensable 
component of the immunological system: 2B4 is a receptor on 
natural killer cells whereas CD48 is expressed on surfaces
of various immune cells. So far, its mechanostability has
not been assessed either experimentally or theoretically.
We find that the dissociation processes 
strongly depend on the direction of pulling and may take place in 
several pathways. Interestingly, the CD48-2B4 interface can be 
divided into three distinct patches that act as units when
resisting the pulling forces. At experimentally accessible pulling speeds, 
the characteristic mechanostability forces are in the range
between 100 and 200~pN, depending on the pulling direction. 
These characteristic forces 
need not be associated with tensile forces involved in the
act of separation of the complex because  prior shear-involving
unraveling within individual proteins may give rise to
a higher force peak. 
%Thus extracting information about adhesion
%forces from pulling experiments may not be straightforward.
\end{abstract}

\vspace*{0.5cm}


\vspace*{0.5cm}

\begin{center}

Proteins: Structure, Function, and Bioinformatics

Accepted Article

DOI: 10.1002/prot.24674

\end{center}

\newpage

\section{Introduction}

A variety of processes in biological cells are performed and
controlled by large protein complexes rather than single proteins.
Examples include replication, transcription, translation, signaling,
protein transport and degradation, and cell adhesion \cite{Alberts}. 
Contemporary structural biology and molecular biophysics are thus 
concerned with understanding of ever larger and more dynamically 
complex molecular assemblies.

Force spectroscopy is a powerful method to study mechanical 
properties of single biopolymers and biomolecular assemblies. 
A typical force spectroscopy experiment uses atomic force microscope (AFM), 
or optical tweezers, to measure the response of a biomolecular system
to stretching, squeezing, or torsional manipulation
\cite{Progress,Rittort,Vogel,Nagy,Crampton,Current}.
These experiments have provided a variety of 
information on transport and force generation in cells,
DNA replication and transcription \cite{Mastrangelo},
DNA unknotting and unwinding \cite{Witz-Stasiak},
ATP synthesis in mitochondria \cite{Scheuring}, 
mechanical behavior of virus capsids
under squeezing \cite{virology,virus1}, and 
other biologically relevant problems. 

There is a growing recent interest in theoretical and experimental
assessments of mechanical stability of protein complexes. If there are
no covalent bonds between individual proteins then stretching 
by the AFM tip may result in dissociation of the complex, 
unraveling of the constituent proteins, or both.
The outcome reflects the strength of the binding forces 
and the structural organization of the complex.
Examples of the systems considered
so far include the titin Z1Z2-telethonin \cite{teleschulten,telerief},
dimeric tubulin \cite{Dima},
the 3$D$ domain-swapped cystatin \cite{cystasikora,cystajaskolski},
oligomeric fibrinogen \cite{Zhmurov},
covalently and non-covalently linked homo-dimeric systems with the cystine knot
motif \cite{dimeric,plugprl}, and the homo-trimeric capsomers of the
CCMV virus capsid \cite{liwo}. 
In particular, Dima and Joshi \cite{Dima} have identified a viscoelastic 
behavior in unraveling of the intermeric couplings of tubulin chains. 
In distinction to the fibrous systems \cite{Zhmurov},
there is no physiologically motivated direction of pulling in the
cystine knot, capsomer, and cystatine complexes 
\cite{dimeric,plugprl,liwo,cystasikora,cystajaskolski}. 
In these systems, the response to stretching 
has been found to be strongly anisotropic: It depends on which site 
the tip is attached to and which site serves as an anchor.
This is a significant extension of the anisotropy
phenomena that have been observed in monomeric proteins \cite{Dietz}.

In this paper, we direct our attention to the
anisotropic effects arising in mechanical manipulation of 
heterodimers. As an illustration we consider
a cell adhesion complex in which mechanical manipulation may lead
either to unfolding, or to dissociation, or to both effects
combined, depending on the geometry of pulling. 
Specifically, we focus on the murine CD48-2B4 complex \cite{Velikovsky_2PPT}
with the Protein Data Bank structure code of 2PTT. 
Protein 2B4 is a heterophilic receptor present at the surface of 
natural killer cells.  The antigenic CD48 proteins 
are expressed on surfaces of various immune cells. 
Interactions of CD48 with 2B4 regulate
the activity of the natural killer cells, which are 
components of the innate immunity against tumors and virally infected cells.
Even though the CD48-2B4 complex appears not to have been studied
through the single molecule manipulation techniques yet, it serves as a
convenient system for theoretical epxlorations.

\begin{figure}[h]
\begin{center}
\scalebox{0.4}{\includegraphics{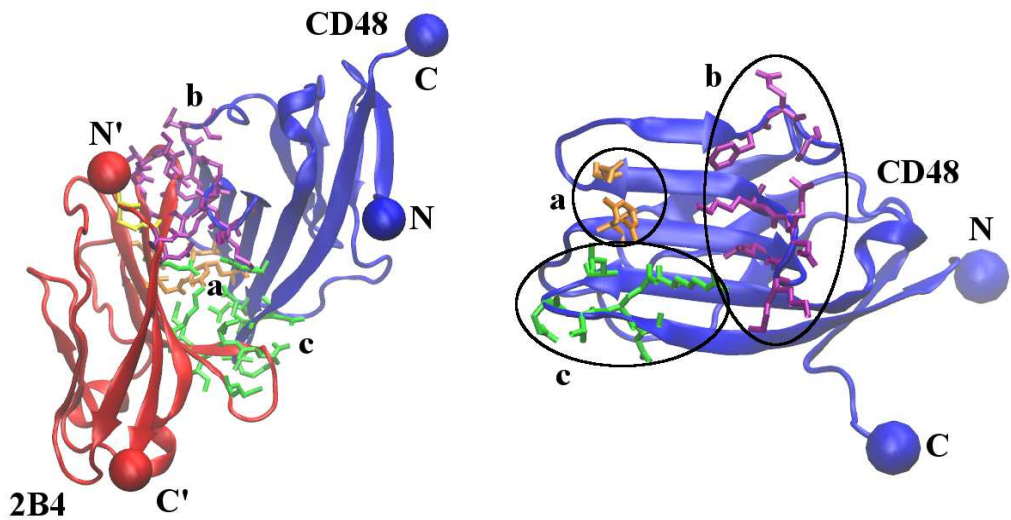}}
\caption{\label{fig:cartoon} 
Left panel: structure of the CD48-2B4 complex. 
Proteins CD48 (chain A) and 2B4 (chain B) are on the right (blue)
and left (red), respectively. 
The residues that are involved in the CD48-2B4 binding 
are shown in the bond representation. 
The pulling simulations reveal three distinct patches of contacts at the 
CD48-2B4 interface, which are denoted by a (orange), b (purple), and c (green).
The disulfide bridge between Cys~3 and Cys~100 in 2B4 is in yellow. 
The termini of CD48 and 2B4 are shown as spheres. 
We denote the CD48 termini by N and C, and the 2B4 termini by N$^{\prime}$ 
and C$^{\prime}$. Right panel: the three patches of the inter-protein contacts
are shown on CD48 on the side that forms the interface with 2B4. }
\end{center}
\end{figure}

The crystal structure of the CD48-2B4 complex that we use in this study 
comprises two extracellular immunoglobulin-like domains:
the C2-type domain of CD48 (shown in blue in Fig.~\ref{fig:cartoon}), 
and the CD48-binding domain of 2B4 (shown in red in Fig.~\ref{fig:cartoon}). 
These domains comprise 110 and 112 amino acid residues, respectively, 
which means that the CD48-2B4 system is too large for a thorough all-atom
molecular dynamics study on relevant time scales. 
Here, we use a coarse-grained structure-based model to study 
the response of the CD48-2B4 complex to stretching at constant speed. 
For simplicity, we consider stretching only by the terminal sites.
Some ways of choosing them may result in partial or complete unfolding of 
the constituent proteins, and other may lead to splitting of 
the complex into separate parts.
We explore the roles of the inter- and intra-protein interactions in 
the competing processes of unfolding and dissociation of the CD48-2B4 complex. 

The 2B4 receptor binds to CD48 with a modest strength, with the
equilibrium dissociation constant in the low micromolar range and 
very fast on and off rates \cite{Velikovsky_2PPT}. In the native state,
19 residues of 2B4 are in contact with 16 residues of CD48, and
most of the contacting residues belong to loops and turns in CD48 and in 2B4.
The CD48-2B4 interface involves only 36 contacts, as determined by atomic 
overlaps (see the Methods section), 
and one of our focal points is quantitative analysis of 
how and when these contacts break during the dissociation process. 

We find that the dissociation scenarios as well as 
the force-displacement patterns depend sensitively on the choice of
the termini that are used to implement stretching. In some cases, 
the proteins separate without any noticeable deformations or structural changes;
in other cases, CD48 unfolds partially before the proteins dissociate.
However, the inter-protein contacts always break in groups rather than sequentially.
The 36 contacts between CD48 and 2B4 can be divided into three distinct patches, 
which are denoted here as a, b, and c, as shown in Fig.~\ref{fig:cartoon}.
We find that the three interface patches cooperate as units 
in resisting the stretching forces.

\begin{figure}[h]
\begin{center}
\scalebox{0.3}{\includegraphics{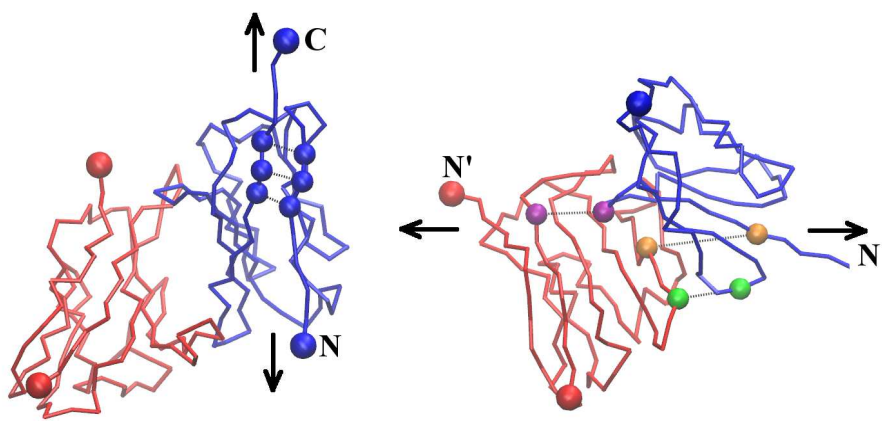}}
\caption{\label{shearvstensil} 
Examples of simulation configurations of the CD48-2B4 complex in which 
shear (left panel) and tensile (right panel) mechanical clamps are formed. 
The arrows show the stretching directions, and the symbols 
(N and C in the left panel; N and N$^{\prime}$ in the right panel) 
lebel the termini to which the stretching forces are applied. 
The black dotted lines mark selected contacts forming the mechanical clamps. 
The residues involved are shown as spheres.
The color code is the same as in Fig.~\ref{fig:cartoon}.  } 
\end{center}
\end{figure}

We also find that a significant resistance to stretching arises due to
two types of mechanical clamps, which involve shear and tensile forces, respectively. 
In shearing, two $\beta$-strands slide by, 
see the left-hand panel of Fig.~\ref{shearvstensil}. 
In tensile strain, the separation between two residues in contact 
increases along the line connecting them, 
as illustrated in the right-hand panel of Fig.~\ref{shearvstensil}. 
Some choices of the pulling termini generate merely tensile effects, 
followed by separation of the protein complex. 
Other choices involve both types of mechanical clamps, 
in which case several force peaks arise on the path to separation. 
The largest of the force peaks needs not be tensile in origin, 
and its height is not necessarily related to the binding strength of the complex.
This observation is relevant for the interpretation of
force spectroscopy experiments on protein complexes.

Since there is no force spectroscopy data on the CD48-2B4 complex 
that we could compare with our simulation results, 
we also simulate stretching of synaptotagmin~1, which 
is a membrane-trafficking multidomain protein that 
has been studied thoroughly in single-molecule pulling experiments 
\cite{Progress,Oberhauser2009,Takahashi2010}. 
Synaptotagmin~1 facilitates fusion of synaptic vesicles. 
Its cytoplasmic part is composed of two domains, 
C2A and C2B, which together form the C2AB complex. 
Our simulation results agree with experimental findings. 
In particular, we find that C2AB is significantly 
less stable than the I27 domain of titin, and that 
the C2A domain in the complex is less stable mechanically 
than its C2B partner. 

\section{Methods}

\subsection{Coarse-grained model \label{section:model}}

We use our coarse-grained structure-based model \cite{Hoang2,JPCM,models,SikoraPLOS2009}, 
in which each amino acid residue is represented 
by a single bead centered on its C$_{\alpha}$ position. 
The beads are tethered together into chains by strong harmonic potentials 
with the spring constant $k_{\rm bond} = 100 \, \epsilon /$\AA$^{2}$, where $\epsilon$ 
is the depth of the potential well associated with the native
contacts, which serves as the basic energy scale in our model. 
The native contacts are identified using an overlap criterion \cite{Tsai} 
applied to the coordinates of all heavy atoms in the native structure. 
However, the amino acid pairs that are very close sequentially, $(i,i+1)$ and $(i,i+2)$, 
are excluded from the contact map \cite{models}. 
It should be noted than our scheme to identify native contacts is different than
in, for example, self-organized polymer model \cite{Dima,Zhmurov} where a uniform
length cutoff is used.

The interactions within the native contacts are described by the Lennard-Jones potential 
\begin{equation}
V^{\rm NAT} (r_{ij}) = 4 \epsilon \left[ \left( \frac{\sigma_{ij}}{r_{ij}} \right)^{12} 
- \left( \frac{\sigma_{ij}}{r_{ij}} \right)^{6}  \right]
\label{eq:6-12}
\end{equation}
Here, $r_{ij}$ is the distance between residues $i$ and $j$ in the contact, 
and the parameters $\sigma_{ij}$ are chosen so 
that each contact in the native structure is 
stabilized at the minimum of the Lennard-Jones potential. 
The value of $\epsilon$ is approximately given by $(110 \pm 30)$~pN~\AA, as
has been determined by comparing simulational results to the experimental
ones on a set of 38 proteins \cite{SikoraPLOS2009}. 
The contacts between the proteins are treated 
in the same manner as the contacts within the proteins 
as both sets are dominated by hydrogen bonds. 
The interactions in the non-native contacts are purely repulsive 
and given by the truncated and shifted Lennard-Jones potential
corresponding to $\sigma_{ij} =  r_0 / \sqrt[6]{2}$ with $r_0 = 4$~{\AA}. 
The energy function comprises also harmonic terms  that favor
the native values of local chiralities in each amino acid chain \cite{Kwiecinska}. 
A harmonic potential with the same force constant
$k_{\rm bond}$ as the C$_{\alpha}$-C$_{\alpha}$ pseudo-bonds 
is used to model the disulfide bond between Cys~3 and Cys~100 in the 2B4 receptor 
(chain B in the PDB structure 2PPT). 

The solvent is implicit and the system evolves in time according 
to the Langevin dynamics. The overall force acting on a particular bead $i$ 
is a sum of three terms: (i) the direct force $\vec{F}_i$ that derives from all the potential 
terms, (ii) the damping force that is proportional to the velocity of the bead, 
and (iii) the random force, $\vec{\Gamma}_i$, that represents thermal noise. 
The corresponding equations of motion 
\begin{equation}
m \frac{ {\rm d}^2 \vec{r}_i }{ {\rm d} t^2} = \vec{F}_i - \gamma \frac{ {\rm d} \vec{r}_i }{ {\rm d} t} + \vec{\Gamma}_i
\end{equation}
are solved by the fifth order predictor-corrector algorithm with the time step of $0.005 \, \tau$.
Here, $\gamma$ is the damping coefficient and all beads are assumed to have the same mass $m$. 
The dispersion of the noise is given by 
$\sqrt{ 2 \gamma k_B T \,}$, where $k_B$ is the Boltzmann constant 
and $T$ denotes the temperature. 
All simulations were performed at $k_B T = 0.3 \, \epsilon$
which is near-optimal in folding kinetics and is of order of the room temperature.
The damping coefficient is set to $\gamma = 2 m / \tau$. This value corresponds
to the overdamped case  -- practically Brownian dynamics -- and
the characteristic time scale, $\tau$,
is of order 1~ns, as argued in Refs.~\cite{Thirumalai,flow}. 

The native contacts can break during the time evolution. Our criterion
for this to happen is that the distance between residues $i$ and $j$ in a contact
exceeds $1.5 \, \sigma_{ij}$.

\subsection{Pulling simulations}

Stretching of the CD48-2B4 complex is implemented by attaching 
harmonic springs to two terminal amino acids. 
One of the springs is fixed in space 
and the other one is moved with a constant speed, $v_p$, 
so that the distance it travels in time $t$ is $d = v_p t$. 
The force constant of the pulling springs is taken as 
$K = 0.12 \epsilon /$\AA$^2$, which corresponds to about 1~pN/nm 
and is close to the elasticity of typical AFM cantilevers \cite{SikoraPLOS2009}. 

\begin{figure}[h]
\begin{center}
\scalebox{0.9}{\includegraphics{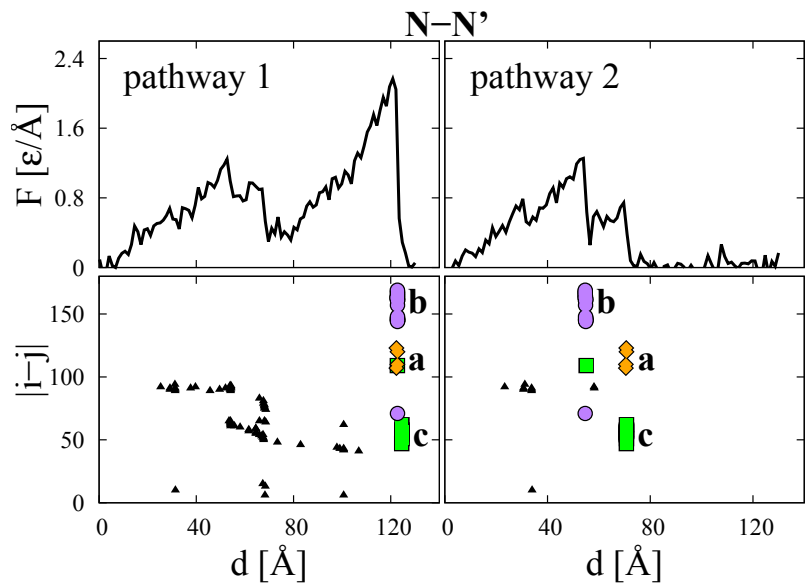}}
\caption{\label{NN'} 
Examples of typical trajectories corresponding
to two pathways for constant speed N-N$^{\prime}$ pulling. 
The top panels show the force-displacement curves. 
The bottom panels show the corresponding
scenario diagrams in which
the intra-protein contacts in CD48 are shown as black triangles whereas 
the contacts between CD48 and 2B4 are shown in orange, purple and green 
to match the three patches a, b and c defined in Fig.~\ref{fig:cartoon}. }
\end{center}
\end{figure}

In our simulations, the response force, $F$, acting on the pulling spring is measured 
and averaged over time periods that correspond to the spring 
displacements of 0.5~{\AA} \cite{SikoraPLOS2009}. 
The $F$-$d$ curves (see the top panels in Fig.~\ref{NN'} and
and several later figures) may come with several peaks, 
and the height of the largest of them will be denoted by $F_{\rm max}$. 
We perform simulations for a range of speeds with the trajectories
evolving for up to 4~ms.
Specific examples of the $F$-$d$ patterns will be shown for 
$v_p=0.002$~{\AA}/$\tau$.

All of the pulling simulations start from the native state. 
In the course of the simulations, the breaking and re-formation of 
native contacts is followed in time. The native contact between 
residues $i$ and $j$ is considered broken if the inter-residue 
distance $r_{ij}$ exceeds a cutoff length, as defined in section~\ref{section:model}.
Due to thermal fluctuations, the broken contacts may get re-established. 
To characterize the unfolding and dissociation pattern of the events, 
we record the spring displacements at which the native contacts 
break for the last time. 
In the corresponding scenario diagrams (see the bottom panels in 
Fig.~\ref{NN'} and several later figures), 
native contacts are labeled by the sequential distance $|j-i|$ 
between the residues involved. Here, we use the convention 
in which the index $i$ runs from 1 to 216 so that the values 
between 1 and 106 correspond to CD48 and between 107 and 216 -- to 2B4. 
Positions of five amino acids are not available in the structure file 
(residues 1, 2 and 3 in chain A, and residues 1 and 110 in chain B) 
and hence the effective sequential size of the system is smaller 
than the nominal 106 plus 110.  The inter-protein contacts in 
patches a, b, and c (shown in Fig.~\ref{fig:cartoon} 
in orange, purple and green, respectively) 
correspond to $i$ around 30 and $j$ around 145, 
$i$ around 40 and $j$ around 195, and finally to $i$ around 90
and $j$ around 150, respectively.

Our model is based on the knowledge of the native structure of 
the protein complex. Both CD48 and 2B4 are $\beta$-proteins, see 
Fig.~\ref{fig:cartoon}. Protein 2B4 comprises two apposing 
$\beta$-sheets. 
One of them is formed by five $\beta$-strands
($\beta_1$, $\beta_8$, $\beta_7$, $\beta_3$ and $\beta_4$) 
and the other one by three $\beta$-strands ($\beta_2$, $\beta_6$ and $\beta_5$). 
Here, the $\beta$-strands are labeled from the N-terminus to the C-terminus. 
2B4 also contains two short helices: $\alpha_1$ located between 
$\beta_4$ and $\beta_5$, and $\alpha_2$ between $\beta_7$ and $\beta_8$.
There are also two $\beta$-sheets in CD48: the first one is formed by 
$\beta_1$, $\beta_9$, $\beta_8$, $\beta_3$, $\beta_4$ and $\beta_5$, whereas
the second one is composed of $\beta_2$, $\beta_7$ and $\beta_6$.
Protein CD48 contains also a short $\alpha$-helix, denoted here by $\alpha_1$, 
which is located between $\beta_7$ and $\beta_8$ in the apposing $\beta$-sheets. 

\section{Results and Discussion}

For any heterodimer, there are six distinct ways to choose a pair of termini, 
and each choice may lead to more than one response pathway in stretching.
Here and below, the termini in CD48 will be denoted by N and C whereas those in 2B4 by
N$^{\prime}$ and C$^{\prime}$, see Fig.~\ref{fig:cartoon}. 

Stretching is applied along the line connecting the two chosen termini. 
To describe the shift in the orientation of stretching in the protocols discussed below, 
we introduce an angle $\theta$ between the C-C$^{\prime}$ direction 
and other directions of stretching. We obtain $\theta$(N-N$^{\prime}$)=68$^{\circ}$, 
$\theta$(N-C$^{\prime}$)=16$^{\circ}$, $\theta$(C-N$^{\prime}$)=23$^{\circ}$,
$\theta$(N-C)=28$^{\circ}$, and $\theta$(N$^{\prime}$-C$^{\prime}$)=38$^{\circ}$. 

\subsection{Dissociation}

We first discuss situations in which the two termini defining the pulling
direction belong to different protein chains. 
This way of pulling results in mechanically induced dissociation
of the two proteins. 

\subsubsection{N-N$^{\prime}$ protocol}

It is useful to first consider the N-N$^{\prime}$ protocol, 
in which the pulling springs are attached to the N-termini of CD48 and 2B4. 
We observe existence of two kinds of dissociation pathways, 
as illustrated in Fig.\ref{NN'}.
In both of them, there is no breaking of contacts within 2B4 because
the disulfide bond between Cys~3 and Cys~100 holds the chain in a sturdy manner. 

On the first pathway, protein CD48 unfolds partially through unzipping 
giving rise to a minor force peak at the spring displacement $d$ of about 55~{\AA}. 
This force peak is associated with breaking of contacts 
between the $\beta_1$-$\beta_2$ and the $\beta_7$-$\alpha_1$ loops, 
and between the $\beta_1$-$\beta_2$ loop and the C-terminal tail of CD48. 
The unzipping process stops at $d$ of about 100~{\AA}, when 
strands $\beta_1$ and $\beta_2$ in CD48 get fully unraveled. 
At $d$ between about 100~{\AA} and 120~{\AA}, 
the protein complex re-orients itself 
so that a tensile clamp between the two proteins is formed,
see the right-hand panel of Fig.~\ref{shearvstensil}. 
Then, at $d$ of about 125~{\AA}, 
all of the interface contacts break almost simultaneously, 
which yields large response forces with $F_{\rm max}$ of 
$2.2 \, \epsilon$/{\AA}, which is comparable to that of the I27 domain
of titin \cite{SikoraPLOS2009}. 

On the second pathway, there is no appreciable unfolding within 
neither of the two units. Instead, the two force peaks 
that are observed are due to rupturing of the
interface contacts. 
The contacts in patch b rupture at $d$ of about 55~{\AA},
which is reflected in moderate response forces 
with $F_{\rm max}$ of about $1.3 \, \epsilon$/{\AA}. 
Then the contacts in patches a and c rupture almost 
simultaneously at $d$ of about 70~{\AA} with a smaller force.
In both rupturing events, we observe tensile mechanical clamps.
What prevents unraveling of CD48 on the second pathway are 
the contacts between $\beta_1$ and $\beta_9$; these contacts break 
early on the first pathway but they never do on the second pathway. 
The first pathway occures more often 
but the level of dominance varies with the speed of pulling, 
as shown in the top panel of Fig.~S1 in Supporting Information (SI). 

\subsubsection{N-C$^{\prime}$ protocol}

\begin{figure}[h]
\begin{center}
\scalebox{0.9}{\includegraphics{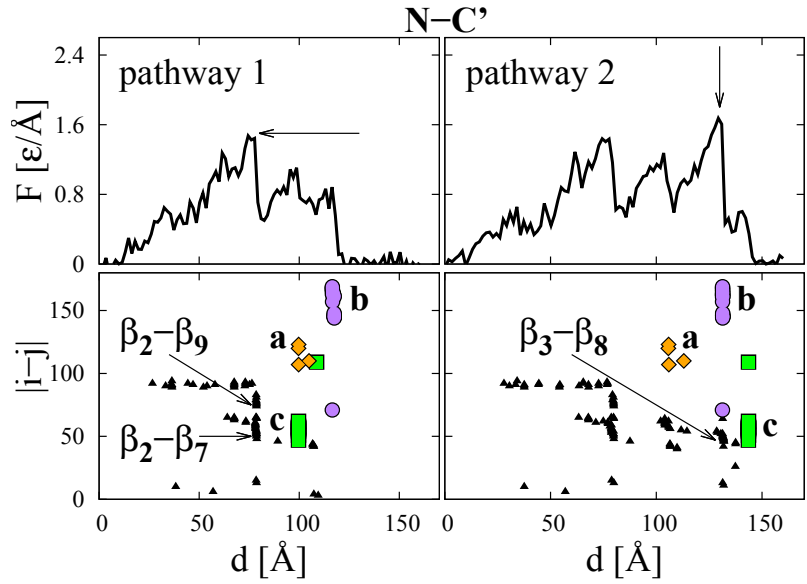}}
\caption{\label{NC'} 
Similar to Fig.~\ref{NN'} but for pulling
in the N-C$^{\prime}$ protocol. The arrows indicate the locations
of the maximal force $F_{\rm max}$. 
}
\end{center}
\end{figure}

Pulling in the N-C$^{\prime}$ protocol also results in two pathways that
lead to separation of the protein complex, see Fig.~\ref{NC'}. 
The two pathways are very similar initially, at $d<80$~{\AA}, 
when $\beta_1$ and $\beta_2$ unravel. 
However, in the first pathway, $F_{\rm max}$
of about 1.6~$\epsilon$/{\AA} arising
at $d \approx 80$~{\AA} is due to shear between 
$\beta_2$ and $\beta_9$, and between $\beta_2$ and $\beta_7$ in CD48, 
as indicated by arrows in Fig.~\ref{NC'}. 
The inter-protein patches break later: first patches a and c 
at $d \approx 100$~{\AA} and then patch b
at $d$ of about 120~{\AA}. The associated force peaks are smaller. 

In the second pathway, unraveling of $\beta_1$ and $\beta_2$ 
gives rise only to a minor force peak at $d \approx 80$~{\AA}. 
The next peak at $d$ of about 110~{\AA} 
is due to rupturing of contacts in patch a. 
However, $F_{\rm max}$ of about 1.8~$\epsilon$/{\AA} 
arises when the contacts in patch b and those 
between $\beta_3$ and $\beta_8$ in CD48
break simultaneously. 
These events occur at $d$ of about 130~{\AA} and 
are immediately followed by the breakage of contacts in patch c
at $d \approx 140$~{\AA}.
The first pathway dominates and its statistical weight 
increases with decreasing speed $v_p$,
see the middle panel of Fig.~S1 in SI. 
In fact, the second pathway is never observed at small 
pulling speeds ($v_p < 0.001$~{\AA}/$\tau$). 

\subsubsection{C-N$^{\prime}$ and C-C$^{\prime}$ protocols}

\begin{figure}[h]
\begin{center}
\scalebox{0.9}{\includegraphics{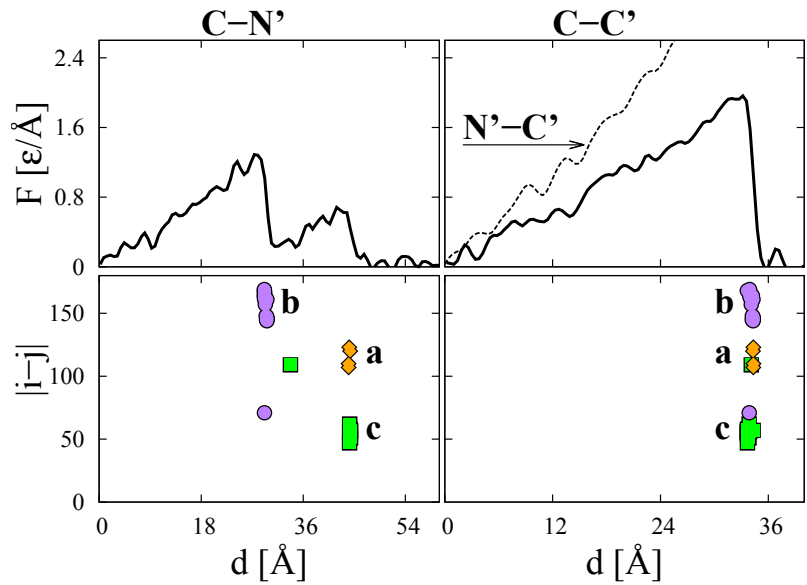}}
\caption{\label{CN'} 
Similar to Fig.~\ref{NN'}  but for the  
C-N$^{\prime}$ (left panels) and
C-C$^{\prime}$ protocols (right panels). There are just single pathways
in these protocols.
}
\end{center}
\end{figure}

In protocols C-N$^{\prime}$ and C-C$^{\prime}$ we observe only
tensile effects and no multiple dissociation pathways. 
The corresponding $F$-$d$ patterns and dissociation scenarios 
are shown in Fig.~\ref{CN'}.
For both protocols, no intra-protein contacts are ruptured before
the two units break loose. In the C-N$^{\prime}$ protocol, 
the contacts in patch b get broken at $d$ of about 30~{\AA}, 
which yields $F_{\rm max}$ of about 1.3~$\epsilon$/{\AA},
and then the contacts in patches a and c break simultaneously 
at $d$ of about 45~{\AA} with smaller forces. 
In the C-C$^{\prime}$ protocol, the three patches break simultaneously 
at $d$ of about 35~{\AA}, and $F_{\rm max}$ increases to 
about 1.9~$\epsilon$/{\AA}. 

\subsection{Unfolding: N-C and N$^{\prime}$-C$^{\prime}$ protocols}

\begin{figure}[h]
\begin{center}
\scalebox{0.9}{\includegraphics{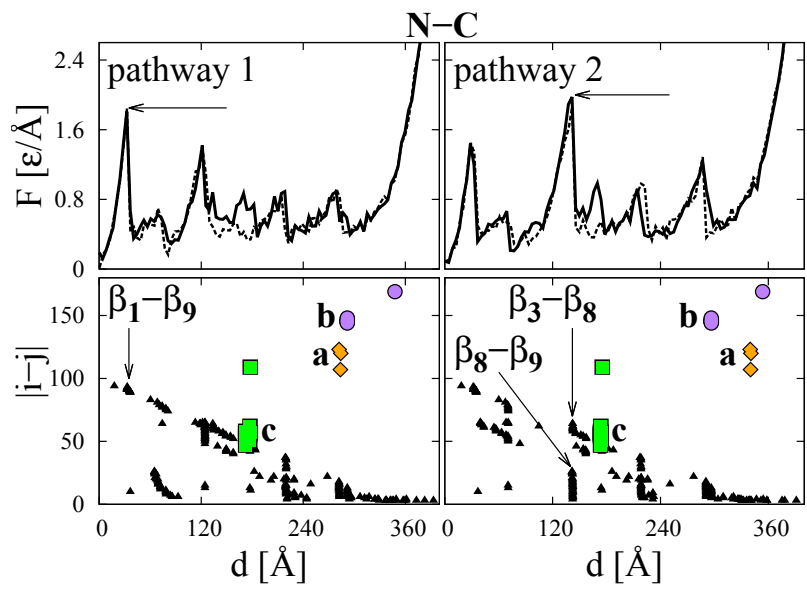}}
\caption{\label{NC} 
Similar to Fig.~\ref{NN'} but for the N-C pulling protocol.
The solid and dashed lines in the $F$-$d$ curves correspond to the presence and 
absence of 2B4, respectively. 
} 
\end{center}
\end{figure}

We now consider situations in which the two termini defining the pulling
direction belong to the same protein chain. 
Pulling in the N$^{\prime}$-C$^{\prime}$ protocol results in a monotonic 
growth of $F$ with $d$ because of the tethering effect 
exerted by the disulfide bond between Cys~3 and Cys~100 in 2B4, 
see the dashed line in Fig.~\ref{CN'}. 
On the other hand, pulling in the N-C protocol results in full 
unraveling of CD48 that can proceed along two pathways. 
The maximum force arises 
either as the first (pathway 1) or the third (pathway 2) 
peak in the $F$-$d$ curves, as indicated by arrows 
in the top panels in Fig.~\ref{NC}. The values of $F_{\rm max}$ are
1.7~$\epsilon$/{\AA} and 1.9~$\epsilon$/{\AA}, respectively.
In the first pathway, the mechanical clamp is due to shearing 
between $\beta_1$ and $\beta_9$ in CD48, 
as illustrated in the left-hand panel of Fig.~\ref{shearvstensil}. 
The shearing between $\beta_1$ and $\beta_9$ occures at $d$ of about 30~{\AA}.
In the second pathway, the mechanical clamp is due to shearing 
between $\beta_8$ and $\beta_9$, and between $\beta_3$ and $\beta_8$,
which takes place at $d \approx 140$~{\AA}.

Interestingly, the larger the sequential distance $|i-j|$ of a contact, 
the earlier its rupture takes place. The contacts in the inter-protein patches get
broken only after the tertiary structure in CD48 is destroyed.
Each of the patches acts coherently, but unlike what happens
in the dissociative protocols, individual patches get affected at
different times.

We observe the same two pathways of CD48 unraveling in the presence and absence 
of the partner protein 2B4, as indicated by the solid and dashed 
lines in Fig.~\ref{NC}, respectively.
Interestingly, as shown in the bottom panel of Fig.~S1, 
the first pathway is statistically more relevant at 
intermediate pulling speeds 
($0.0002$~{\AA}$/\tau < v_p < 0.005$~{\AA}$/\tau$),
and the second pathway dominates at low and high 
pulling speeds. 

\subsection{Shear and tensile mechanical clamps}

\begin{figure}[h]
\begin{center}
\scalebox{1.0}{\includegraphics{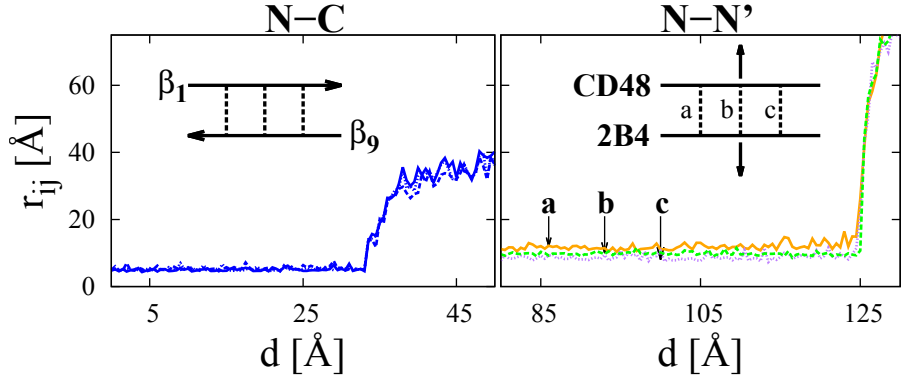}}
\caption{\label{mech_clamps} 
Inter-residue distances $r_{ij}$ as a function of the spring displacement $d$. 
Left panel: Distances $r_{ij}$ between residues $i=6$ and $j=97$, $i=7$ and $j=98$, and $i=8$ and $j=98$, 
which represent three exemplary contacts between strands $\beta_1$ and $\beta_9$ in CD48. 
These residues are shown as blue spheres in the left-hand panel of Fig.~\ref{shearvstensil}. 
The moderate increase at $d \approx 32$~{\AA} corresponds to shear between $\beta_1$ and $\beta_9$. 
Right panel: Distances $r_{ij}$ between residues $i=28$ and $j=148$, 
$i=35$ and $j=203$, and $i=88$ and $j=145$, 
which represent three selected contacts between CD48 and 2B4 in patches a, b and c, respectively. 
These contacts are shown in the right-hand panel of Fig.~\ref{shearvstensil}. 
The steep increase at $d \approx 125$~{\AA} indicates action of the tensile mechanical clamps. 
} 
\end{center}
\end{figure}

As discussed before, $F_{\rm max}$ arises either as a result of shear 
between $\beta$-strands in CD48 or due to the tensile clamps 
between CD48 and 2B4, depending on the pulling protocol and pathway. 
The distinction between the two kinds of mechanical clamps 
sits in the directionality of the movements of the residues involved,
as illustrated by the simulation snapshots in Fig.~\ref{shearvstensil}. 
However, it also shows in the time dependence of 
the distances between the residues in contacts that form the clamp. 
The left-hand panel of Fig.~\ref{mech_clamps} shows three such distances 
as a function of $d$ in pathway 1 in the N-C protocol.
These distances correspond to three native contacts between 
$\beta_1$ and $\beta_9$ in CD48.
For $d<32$~{\AA}, the inter-residue distances $r_{ij}$ 
fluctuate around their native values. At a critical distance of about 32~{\AA}, 
shearing between the two $\beta$-strands leads to the simultaneous breaking of 
the three contacts, which is reflected in the fast increase in the inter-residue 
distances $r_{ij}$ from about 5~{\AA} to about 35~{\AA}. 
In the right-hand panel of Fig.~\ref{mech_clamps}, 
the three distances $r_{ij}$ correspond to contacts 
between CD48 and 2B4 in patches a, b and c, respectively. 
The situation considered here relates to pathway 1 in the N-N$^{\prime}$ protocol.
For $d<125$~{\AA}, the inter-residue distances $r_{ij}$ 
fluctuate around their native values. At a critical distance $d \approx 125$~{\AA},
the contacts in patches a, b and c break simultaneously, which leads to 
a sudden increase in the distances $r_{ij}$ from about 10~{\AA} to about 70~{\AA}.
The difference between the two clamps is in the steepness and magnitude of the increase of 
the inter-residue distances -- from 5 to 35~{\AA}, and from 10 to 70~{\AA} 
in the case of shear and tensile clamps, respectively.
The simulation snapshots in Fig.~\ref{shearvstensil} correspond, 
left-to-right, to the panels of Fig.~\ref{mech_clamps}.

\subsection{The velocity dependence}

\begin{figure}[h]
\begin{center}
\scalebox{0.9}{\includegraphics{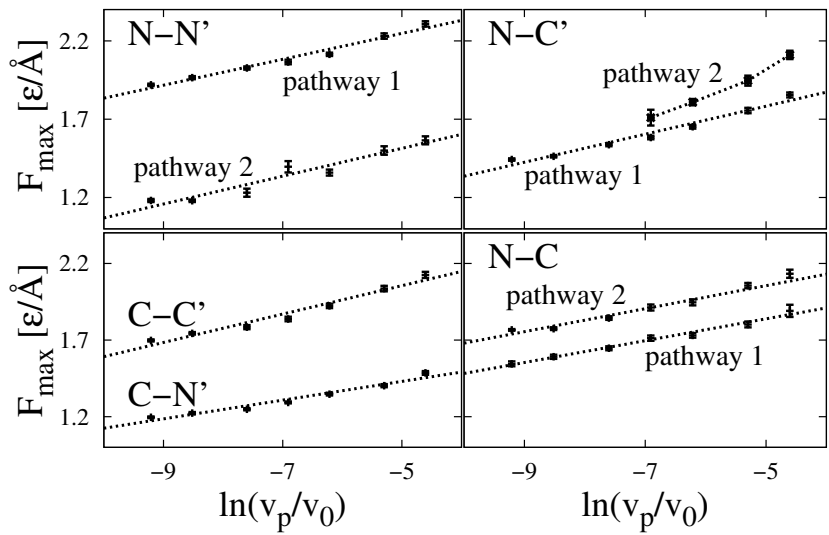}}
\caption{\label{vel} 
$F_{\rm max}$ as a function of $v_p$
in the logarithmic scale for the indicated pulling protocols and pathways. 
The unit of speed is $v_0$=1~{\AA}/$\tau$. 
The dotted lines correspond to the relation $F_{\rm max} = p \log (v_p / v_0) + q$ 
with the fitting parameters $p$ and $q$ as specified in Table~\ref{tab:pq}. 
}
\end{center}
\end{figure}

Our discussion so far has been focused mainly on trajectories 
obtained at one particular speed of pulling, $v_p = 0.002$~{\AA}/$\tau$. 
We now discuss how $F_{\rm max}$ varies with $v_p$, see Fig.~\ref{vel}. 
We performed simulations at seven different speeds of pulling 
ranging from $10^{-4}$\AA/$\tau$ to $10^{-2}$\AA/$\tau$, which corresponds
to the experimental speeds between 10~nm/ms and 1~nm/$\mu$s. 
At each of the pulling speeds we ran at least 60 trajectories. 
We separated the simulation data into pathways, if any, 
and averaged over the trajectories. 
We find an efective logarithmic dependence \cite{Bell,Evans1997}
\begin{equation}
F_{\rm max} = p \ln \left( v_p / v_0 \right) + q
\label{F_max_eq}
\end{equation}
in the range of the speeds studied. 
Here, the unit of speed is $v_0$=1~{\AA}/$\tau$. 
The fitting parameters $p$ and $q$ are summarized in Table~\ref{tab:pq}. 
The parameter $p$ describes how fast $F_{\rm max}$ changes with $v_p$.
The range of variations in $p$ is seen to be within a factor of 1.5, 
which reflects anisotropy effects. 
Interestingly, the two pathways of the N-C protocol exhibit almost identical values 
of the parameter $p$. 
The slowest dependence is observed for the C-N$^{\prime}$ protocol whereas
the most rapid one is seen for the second pathway of the N-N$^{\prime}$ protocol. 

Assuming the Bell-Evans model \cite{Bell,Evans1997} in which 
\begin{equation}
F_{\rm max} = \frac{k_{B}T}{x^\ddagger} 
\ln \left( \frac{x^\ddagger K v_p}{k_{B} T k_0} \right)
\label{Bell}
\end{equation}
one can extract the intrinsic off-rates of the interacting proteins, $k_0$, 
and the location of the free-energy barriers confining the proteins 
in their bound state, $x^\ddagger$, from the fitting parameters $p$ and $q$. 
Comparison of Eqs.~(\ref{F_max_eq}) and (\ref{Bell}) gives
$x^\ddagger = k_{B}T/p$ and $k_0 = e^{-q/p} K v_0 / p$. 
The resulting values of $k_0$ and $x^\ddagger$ are summarized in Table~\ref{tab:pq}. 
In case of deviations from the logarithmic dependence (\ref{Bell}), 
one can use more sophisticated theories \cite{Dudko2006} 
that permit to extract also the height of the free-energy 
barriers $\Delta G^\ddagger$. 

It is interesting to observe that the smallest off-rates 
in Table~\ref{tab:pq} correspond to either complete (protocol N-C) 
or partial (N-N$^{\prime}$ pathway 1) unfolding of CD48. 
The off-rates that correspond merely to dissociation (protocols C-C$^{\prime}$, 
C-N$^{\prime}$, and N-N$^{\prime}$ pathway 2) are orders of magnitude larger. 
Nevertheless, some values of $k_0$ as given in Table~\ref{tab:pq} seem
unreasonably small.
They have been obtained at $T = 0.3 \, \epsilon/k_B$.
Higher temperatures should lead to their substantial increase.
Note that the room temperature is more likely to correspond
to about $0.35 \, \epsilon/k_B$ (notice that the value of $\epsilon$ 
itself comes with about 25~\% uncertainty \cite{SikoraPLOS2009}) 
and the physiological temperatures are still higher. 

We note that when the pulling sites are the N- and C$^{\prime}$-termini, 
the second pathway occurs only at the pulling speeds $v_p \ge 0.001$~\AA/$\tau$, 
which is indicated by the absence of data points at small speeds $v_p < 0.001$~\AA/$\tau$. 
To further investigate this observation, we calculated the frequency of 
occurrence of the first pathway in protocols N-N$^{\prime}$, N-C$^{\prime}$ and N-C.
At a particular speed of pulling, the frequency $\nu_1$ is defined as the ratio of 
the number of trajectories at which the first pathway has occured to 
the total number of trajectories. 
The occurrence frequency $\nu_1$ as a function of the pulling speed $v_p$ is shown in Fig.~S1. 
In the N-N$^{\prime}$ protocol, the frequency $\nu_1$ varies between 0.6 and 0.95, and 
attains the maximum value at $v_p = 10^{-3}$~{\AA}/$\tau$. 
In the N-C$^{\prime}$ protocol, $\nu_1$ increases from 0.65 at $v_p = 10^{-2}$~{\AA}/$\tau$ to
1 at $v_p = 10^{-3}$~{\AA}/$\tau$, and remains constant, $\nu_1 = 1$, in the interval between 
$v_p = 10^{-3}$~{\AA}/$\tau$ and $v_p = 10^{-4}$~{\AA}/$\tau$. 
In the N-C protocol, the frequency $\nu_1$ increases with the speed $v_p$ in the interval 
between $v_p = 10^{-4}$~{\AA}/$\tau$ and $v_p = 10^{-3}$~{\AA}/$\tau$, and decreases with $v_p$ 
in the interval from $v_p = 10^{-3}$~{\AA}/$\tau$ to $v_p = 10^{-2}$~{\AA}/$\tau$, 
with the maximal value of 0.7 at $v_p = 10^{-3}$~{\AA}/$\tau$. 
The error bars in Fig.~S1 indicate the standard deviation of the Bernoulli process. 

We now turn our attention to the process of CD48-2B4 separation. 
The separation event takes place at the moment when no contacts are present 
between CD48 and 2B4 for the first time on the trajectory. 
The corresponding spring displacement at which separation takes place 
will be denoted by $d_s$. Fig.~S2 in SI shows that $d_s$ 
decreases with $v_p$ in a way that depends on the protocol and the pathway. 
Its most significant dependence is observed for pathway 2 
in protocal N-C$^{\prime}$, which is the pathway 
that finds no continuation at small speeds.
The decrease in $d_s$ reflects the greater role of thermal
fluctuations at lower speeds since fluctuations foster separation.

The spring displacement $d$ does not directly relate to the end-to-end distance, $L$,
between the pulling termini. In fact, $L$ reflects transformations in 
the structure in a more direct manner. Fig.~S3 in SI shows 
that $L_s$ -- i.e. the value of $L$ corresponding to separation -- also depends 
on $v_p$, although usually less sensitively. For instance, there is much less
variation in $L_s$ compared to $d_s$ in the C-C$^{\prime}$ protocol.
On the other hand, they are about the same for both pathways in the
N-C$^{\prime}$ protocol. 

\subsection{Stretching of the multidomain synaptotagmin~1 }

To our knowledge, there have been no single-molecule pulling 
experiments performed on the CD48-2B4 complex. 
Thus the assumptions underlying the construction of our model, 
especially about the relative strength of the inter-protein 
contacts remain untested. 
To establish connections to force spectroscopy experiments on 
membrane-associated multidomain proteins, 
we performed pulling simulations of synaptotagmin~1. 
The latter protein is involved in remodeling 
the plasma membrane during neurotransmitter release at the synapse. 
The cytoplasmic region of synaptotagmin~1 contains two domains, 
C2A and C2B, which together form the C2AB module. These two domains 
are quite similar structurally but have been found to exhibit 
different mechanical stabilities within the C2AB module 
\cite{Oberhauser2009,Dima2011}. 

\begin{figure}[h]
\begin{center}
\scalebox{1.0}{\includegraphics{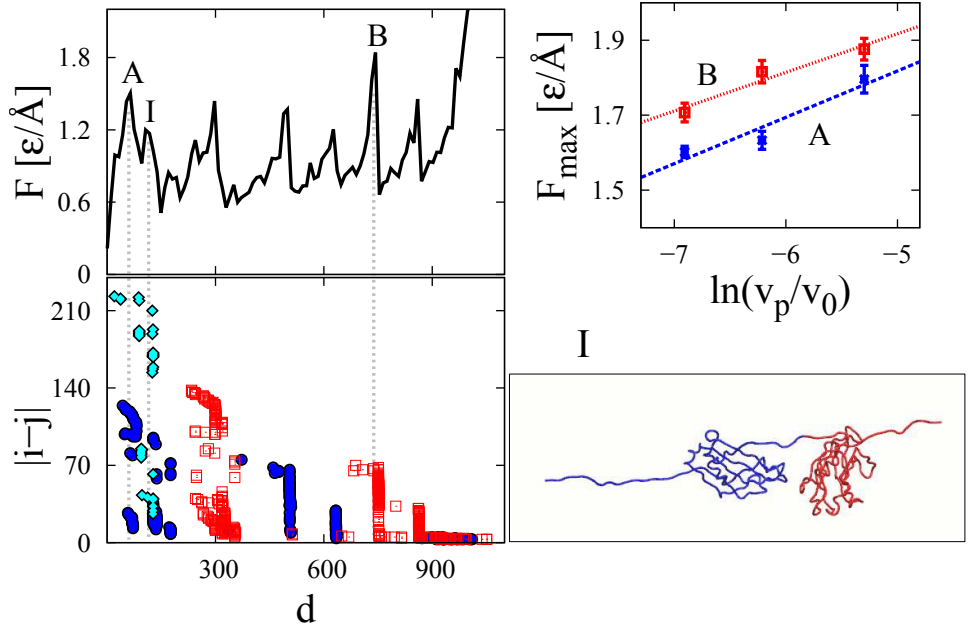}}
\caption{\label{synaptotagmin1} 
Pulling simulations of synaptotagmin 1. 
In the scenario diagram, the contacts within the separate 
C2A and C2B domains are shown as filled (blue) circles and 
empty (red) squares, respectively. 
The inter-domain contacts are represented by filled (cyan) diamonds. 
The highest peak in the $F$-$d$ curve 
at $d \approx 740$~{\AA} is denoted by B. 
It is associated with rupturing the core of the C2B domain. 
The second highest peak at $d \approx 60$~{\AA} is denoted by A. 
It is associated with unraveling of 
$\beta_1$ and $\beta_2$ in the C2A domain. 
The minor force peak that is related to breaking the C2AB interface 
at $d \approx 120$~{\AA} is denoted by I. 
A typical simulation configuration at stage I is 
shown in the lower right corner. 
The dependences of $F_{\rm max}$ on $v_p$ for the force peaks A and B 
are shown in the upper right panel 
(data averaged over a dozen trajectories at each of the pulling speeds). }
\end{center}
\end{figure}

The issue which is analogous to that pertaining to the 
CD48-2B4 complex is whether it is indeed sensible to 
assume that the inter-domain contact energy, which we 
denote here by $\epsilon^{\prime}$, is about the same 
as the contact energy $\epsilon$ within each domain. 
Fig.~\ref{synaptotagmin1} shows a typical $F$-$d$ curve and 
the corresponding scenario diagram for $\epsilon^{\prime} = \epsilon$
and $v_p = 0.002$~{\AA}/$\tau$.
The highest force peak (labelled as B in Fig.~\ref{synaptotagmin1}) 
is associated with rupturing the core of the C2B domain. The second highest 
peak (labeled as A in Fig.~\ref{synaptotagmin1}) arises when  
$\beta_1$ and $\beta_2$ of the C2A domain unravel. This event is followed by 
breaking the C2AB interface, which results in a minor force peak 
(labeled by I in Fig.~\ref{synaptotagmin1}). A typical simulation 
configuration at this stage is depicted in the lower right corner of 
Fig.~\ref{synaptotagmin1}. It is similar to the relevant 
configuration obtained in steered molecular dynamics simulations 
performed at a pulling speed that is three orders of magnitude 
larger than used here \cite{Oberhauser2009}. 

After extrapolation to the experimental pulling speed, see 
Fig.~\ref{synaptotagmin1}, we obtain the values of $F_{\rm max}$ 
to be equal to $(100 \pm 30)$~pN and $(70 \pm 20)$~pN for 
the force peaks B and A, respectively. These values are consistent 
with the experimental results of about 100~pN and 50~pN 
(with the statistical error of about 20~pN) 
for the C2B and C2A domains \cite{Oberhauser2009}. 

We now consider the situation in which $\epsilon^{\prime}$ is distinct 
from $\epsilon$. We find that if $\epsilon^{\prime} = 2 \epsilon$
then the C2A domain becomes much more stable than the C2B domain,
which disagrees with experimental findings. 
On the other hand, if $\epsilon^{\prime} = \epsilon / 2$
then the second highest force peak becomes
associated with rupturing the core of the C2A domain rather than
with unraveling of $\beta_1$ and $\beta_2$, see
Fig.~S4 in SI.  However, the characteristic
forces ($1.6 \, \epsilon$/{\AA} and $1.75 \, \epsilon$/{\AA}
for C2A and C2B, respectively, at $v_p = 0.002$~{\AA}/$\tau$)
are very close to those presented in Fig.~\ref{synaptotagmin1}
($F_{\rm max}^{\rm A} = 1.6 \, \epsilon$/{\AA} and
$F_{\rm max}^{\rm B} = 1.8 \, \epsilon$/{\AA}) at the same speed of pulling.
Nevertheless, the appearance of the whole $F$-$d$ pattern is much
closer to the experimental one when $\epsilon^{\prime} = \epsilon$ than when
$\epsilon^{\prime}$ is reduced.
We therefore conclude that assuming the equality of $\epsilon^{\prime}$ 
and $\epsilon$ is consistent with the experimental results and with
the similarity of the physical nature of the interdomain contacts. 

\section{Conclusions}

The mechanical stability of adhesion proteins and their complexes is necessary 
to sustain interactions and signaling between cells. 
To investigate the mechanical stability of the adhesion protein complex CD48-2B4, 
we have performed pulling simulations using our coarse-grained structure-based model. 
We find that the force peaks arise either as a result of shear 
between $\beta$-strands in CD48  or due to tensile mechanical clamps 
between CD48 and 2B4, depending on the mode of pulling and 
dissociation pathway.  If there are several force peaks in a given
$F(d)$ curve then, in general, the maximal one may be associated with shear 
or with the tensile separation. In the former case, the measurement of
$F_{\rm max}$ would not provide an estimate of the adhesion forces.

\begin{figure}[h]
\begin{center}
\scalebox{0.7}{\includegraphics{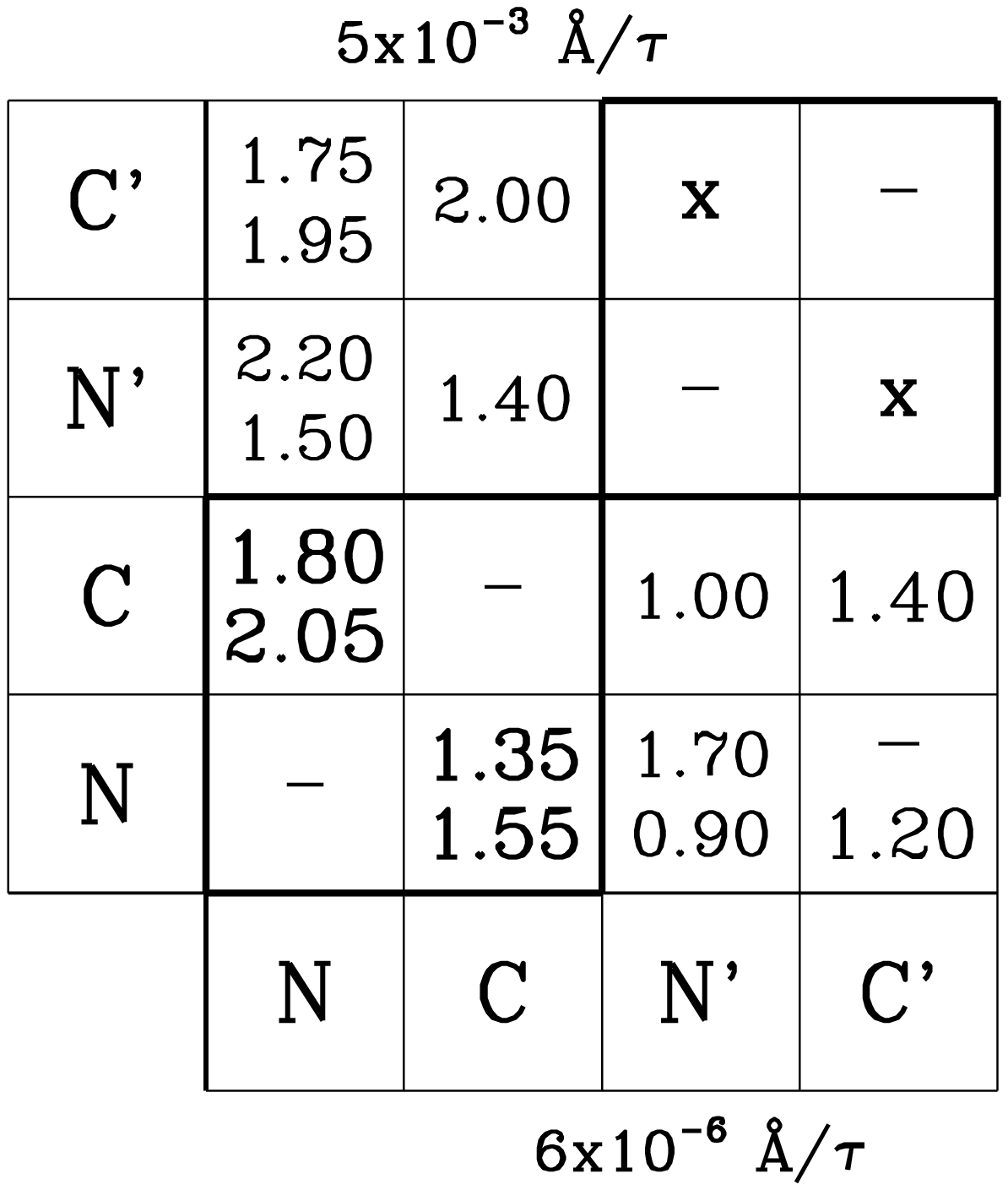}}
\caption{\label{table} 
Values of $F_{\rm max}$ for the
various protocols of pulling. The data above the diagonal correspond to 
$v_p = 5 \times 10^{-3}$~{\AA}/$\tau$. Those below the diagonal -- to 
the pulling speed  of $6 \times 10^{-6}$~{\AA}/$\tau$ 
as obtained through extrapolation. 
A double entry indicates existence of two pathways. 
The entries just next to the diagonal
are highlighted as they correspond to same-chain pulling.
} 
\end{center}
\end{figure}

Table \ref{table} provides a summary of our findings regarding the values of
$F_{\rm max}$ for various protocols of stretching. 
For a given $v_p$, the table is symmetric. We present it in a way which
gives values of $F_{\rm max}$ for two pulling speeds: above the diagonal 
the data correspond to $v_p$ of $5 \times 10^{-3}$~{\AA}/$\tau$ -- the standard
used in our surveys of mechanostability \cite{SikoraPLOS2009} -- and
below the diagonal they correspond to $6 \times 10^{-6}$~{\AA}/$\tau$,
which is in the lower range of speeds used in protein-related
AFM experiments \cite{SikoraPLOS2009}. The highlighted diagonal blocks in 
the table correspond to same-chain pulling which results either in
structure unfolding (N-C) or just deformation (N$^{\prime}$-C$^{\prime}$).
The off-diagonal blocks correspond to pulling of different chains
which results in dissociation. The $v_p$ dependencies are seen to be
comparable in these two classes of pullings, and dissociation may
come either with a smaller or larger $F_{\rm max}$ than unraveling.

We observe existence of strong anisotropies. For dissociation,
$F_{\rm max}$ varies between 0.9 and 1.7 $\epsilon$/{\AA} at the
extrapolated experimental speed. For unraveling -- between 1.35 and
1.55 $\epsilon$/{\AA}.
In the N-C protocol, and also on the first pathway of the N-C$^{\prime}$ protocol,
$F_{\rm max}$ is caused by shear between $\beta$-strands in CD48.
In other pulling protocols, $F_{\rm max}$ is due to tensile mechanical clamps 
between CD48 and 2B4. It is important to note that the largest force signal
during dissociation may actually come from shearing and not tensile separation.

Both CD48 and 2B4 are anchored in membranes through the C-terminal ends \cite{Velikovsky_2PPT}. 
The forces acting on the CD48-2B4 complex during cell adhesion are thus applied 
in the C-C$^{\prime}$ direction (although membrane fluctuations on nanometer scales 
may alter this direction to some extent \cite{HuWeikl_PNAS_2013}). 
Interestingly, we find that in the C-C$^{\prime}$ 
pulling simulations neither CD48 nor 2B4 unravels (both proteins maintain their structure) 
and the entire CD48-2B4 interface acts as a whole to resist external forces with 
$F_{\rm max} = 1.4$~$\epsilon$/{\AA} at the experimental speed of pulling. 
The latter value corresponds to about 150~pN. 
These results suggest that the CD48-2B4 complex is adapted 
to resist significant forces while maintaining the structure of 
individual proteins. In particular, the disulfide bond between Cys~3 and Cys~100 
prevents the rupturing of contacts within 2B4.

\clearpage

\noindent {\bf Acknowledgments}

\noindent  This work has been supported by the Polish National Science Centre Grants 
No. 2012/05/B/NZ1/00631 (BR) and 2011/01/B/ST3/02190 (MC) as well as
by the ERA-NET Grant FiberFuel (MC). 
{\L}M acknowledges access to the KRUK computer cluster at the Faculty of Physics,
University of Warsaw.

\clearpage

\section*{Tables and figures}

\begin{table}[h]
\begin{center}
\begin{tabular}{|l|l|l|l|l|l|}
\hline 
protocol and pathway & $p$ [$\epsilon$/{\AA}] & $q$ [$\epsilon$/{\AA}] & $x^\ddagger$ [{\AA}] & $k_0$ [1/ $\tau$] &\# trajectories \\
\hline
N-N$^{\prime}$ pathway 1 & 0.083 & 2.66 & 3.6 & $2 \cdot 10^{-14}$ & 373 \\
\hline
N-N$^{\prime}$ pathway 2 & 0.089 & 1.96 & 3.4 & $4 \cdot 10^{-10}$ & 92 \\
\hline
N-C$^{\prime}$ pathway 1 & 0.089 & 2.23 & 3.4 & $2 \cdot 10^{-11}$ & 398 \\
\hline
N-C$^{\prime}$ pathway 2 &   N/A  & N/A  & N/A & N/A & 67 \\
\hline
C-N$^{\prime}$ & 0.061 & 1.74 & 4.9 & $9 \cdot 10^{-13} $ & 420 \\
\hline
C-C$^{\prime}$ & 0.093 & 2.52 & 3.2 & $2 \cdot 10^{-12} $ & 420 \\
\hline
N-C pathway 1 & 0.071 & 2.20 & 4.2 & $7 \cdot 10^{-14}$ & 231 \\
\hline
N-C pathway 2 & 0.075 & 2.43 & 4.0 & $2 \cdot 10^{-14}$ & 239 \\
\hline
\end{tabular} 
\caption{ 
The fitting parameters $p$ and $q$ in Eq.~(\ref{F_max_eq}), 
and the resulting parameters $k_0$ and $x^\ddagger$ in Eq.~(\ref{Bell}),
for the various pulling protocols and pathways. 
The differences in $p$ and $q$ between the protocols reflect the anisotropic 
response of the system to pulling. }
\label{tab:pq}
\end{center}
\end{table}

\clearpage

\section*{ Supporting Information }

\setcounter{figure}{0}
\makeatletter 
\renewcommand{\thefigure}{S\@arabic\c@figure}
\makeatother

This section contains four figures that support results reported in
{\sl "Unbinding and unfolding of adhesion protein complexes through stretching:
interplay between shear and tensile mechanical clamps"} 
by Bartosz R\'{o}\.{z}ycki, {\L}ukasz Mioduszewski and Marek Cieplak

\vspace*{0.5cm}

\begin{figure}[h]
\begin{center}
\scalebox{0.9}{\includegraphics{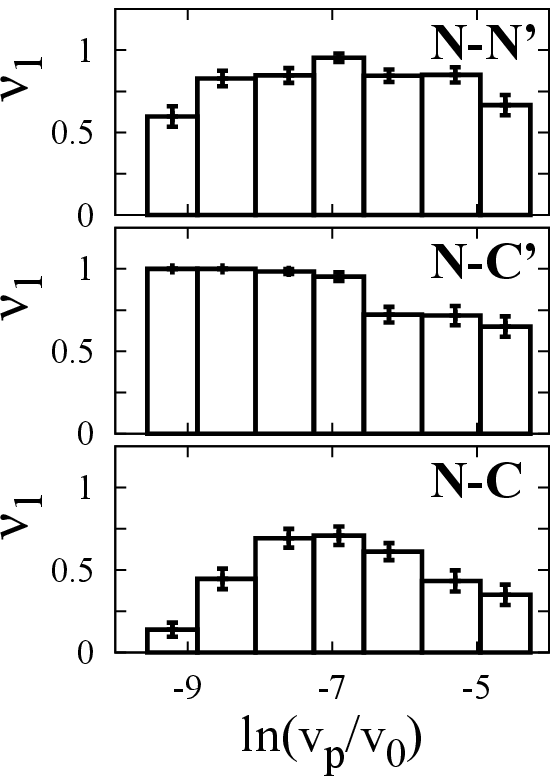}}
\caption{ \label{fig:w_v}
The relative frequency of appearance of the first pathway, 
$\nu_1$, as a function of the pulling speed $v_p$  for the
N-N$^{\prime}$, N-C$^{\prime}$ and N-C protocols as indicated. 
}
\end{center}
\end{figure}

\begin{figure}[h]
\begin{center}
\scalebox{0.8}{\includegraphics{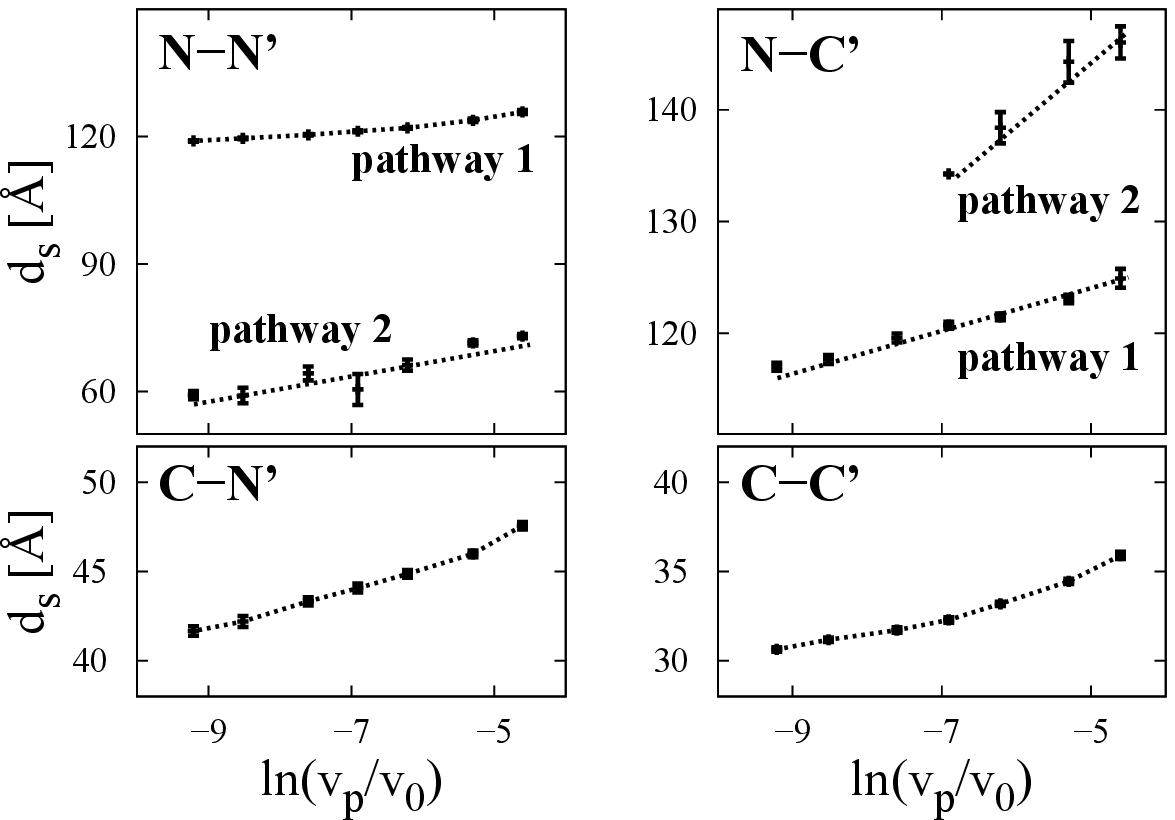}}
\caption{ \label{fig:dmax_v} 
Separation distance $d_s$ 
as a function of the pulling speed $v_p$ 
in the logarithmic scale for the indicated pulling protocols and pathways. 
The dotted lines are shown to guide the eye. 
}
\end{center}
\end{figure}

\begin{figure}[h]
\begin{center}
\scalebox{0.8}{\includegraphics{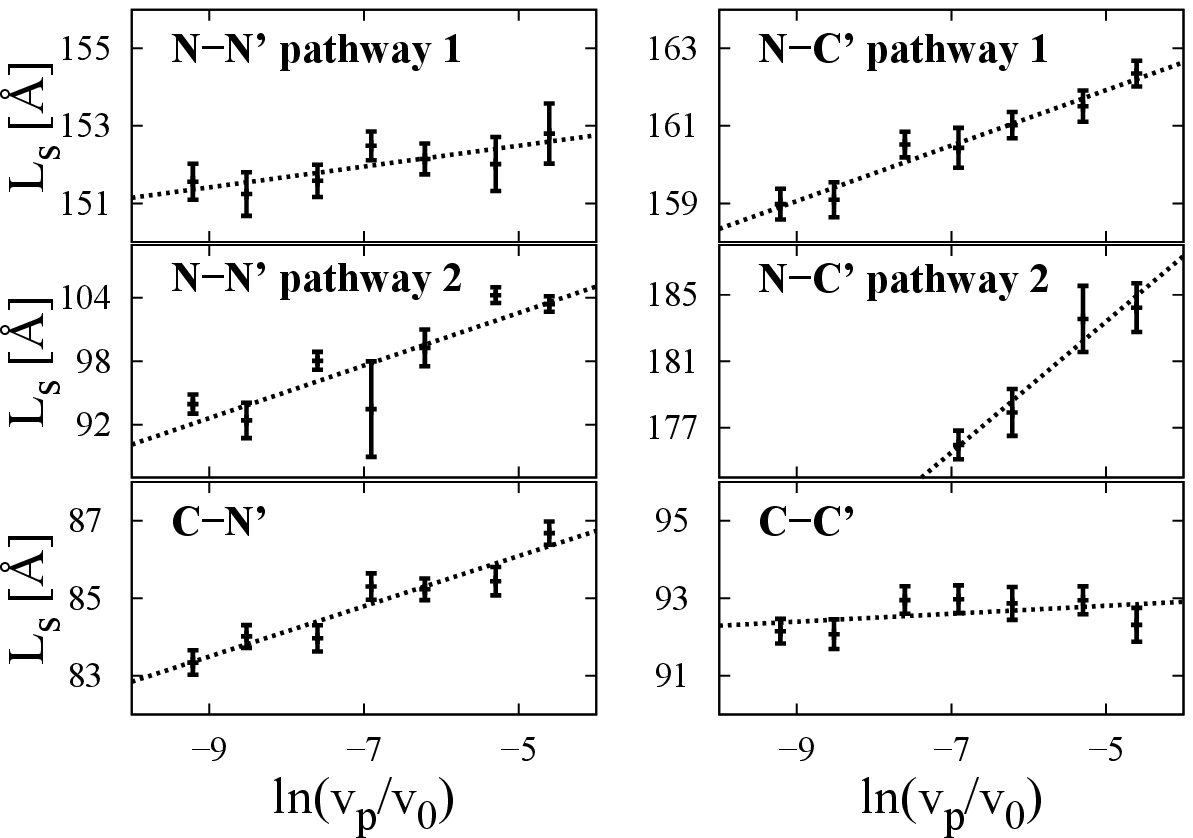}}
\caption{ \label{fig:Lmax_v} 
The end-to-end distance at the time of 
CD48-2B4 separation, $L_s$, as a function of the pulling speed $v_p$ 
in the logarithmic scale. 
The dotted lines are drawn to guide the eye.
}
\end{center}
\end{figure}

\begin{figure}[h]
\begin{center}
\scalebox{0.8}{\includegraphics{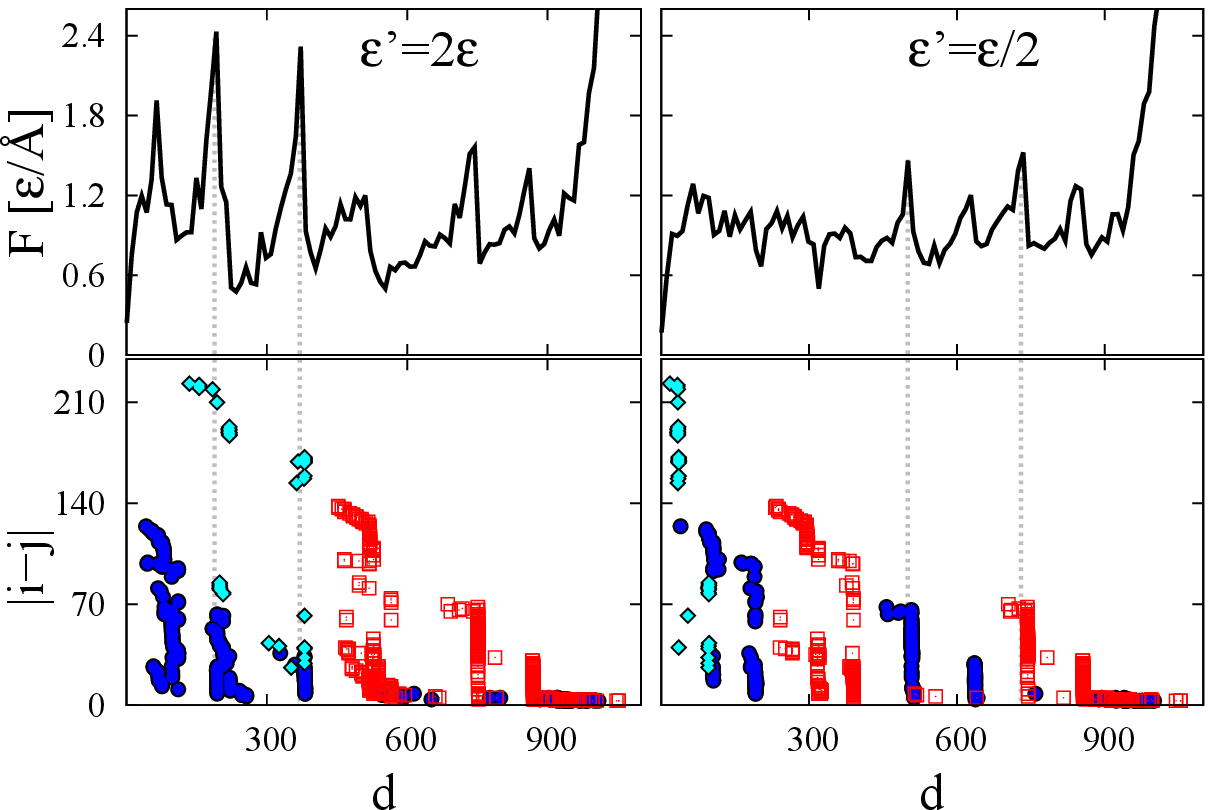}}
\caption{ \label{synaptotagmin2}
Pulling simulations of synaptotagmin 1 with the inter-domain 
contact energies $\epsilon^{\prime} = 2 \, \epsilon$ (left) and
$\epsilon^{\prime} = \epsilon / 2$ (right) 
at $v_p = 0.002$~{\AA}/$\tau$. 
The top panels show typical $F$-$d$ curves. The bottom panels 
show the corresponding scenario diagrams. Here, the contacts 
within the C2A domain are shown as filled (blue) circles, 
those within the C2B domain as empty (red) squares, 
and the inter-domain contacts as filled (cyan) diamonds. 
The two highest force peaks are indicated by dashed lines. }
\end{center}
\end{figure}


\vspace*{0.5cm}


\begin{thebibliography}{90}

\bibitem{Alberts}
Alberts B, Johnson A, Lewis J, Raff M, Roberts K, Walter P. 
{\it Molecular Biology of the Cell}. New York: Garland Science; 2002.

\bibitem{Progress}
Carrion-Vazquez M, Oberhauser AF, Fisher TE, Marszalek PE, Li H, Fernandez JM. 
{\sl Mechanical design of proteins studied by single-molecule force
spectroscopy and protein engineering}. 
Prog Biophys Mol Biol 2000;74:63-91.

\bibitem{Rittort}
Rittort F. {\sl Single molecules experiments in biological physics:
methods and applications}. J Phys Cond Mat 2006;18:531-583.

\bibitem{Vogel}
Vogel V. {\sl Mechanotransduction involving multimodular proteins: converting force into
biochemical signals}. Annu Rev Biophys Biomol Struct 2006;35:459-488.

\bibitem{Nagy}
Nauman KC, Nagy A. {\sl Single-molecule force spectroscopy: 
optical tweezers, magnetic tweezers and atomic force microscopy}. 
Nat Meth 2008;5:491-505.

\bibitem{Crampton}
Crampton N, Brockwell DJ. 
{\sl Unravelling the design principles for single protein mechanical strength}. 
Curr Opin Struct Biol 2010;20:508-517.

\bibitem{Current}
Galera-Prat A, Gomez-Sicilia A, Oberhauser AF, Cieplak M, Carrion-Vazquez M. 
{\sl Understanding biology by stretching proteins: recent progress}. 
Curr Opin Struct Biol 2010;20:63-69.

\bibitem{Witz-Stasiak}
Witz G, Stasiak A.
{\sl DNA supercoiling and its role in DNA decatenation and unknotting}.
Nucl Acids Res 2010;38:2119-2133.

\bibitem{Scheuring}
Buzhynskyy N, Sens P, Prima V, Sturgis JN, Scheuring S. 
{\sl Rows of ATP Synthase Dimers in Native Mitochondrial Inner Membranes}. 
Biophys J 2007;93:2870-2876.

\bibitem{Mastrangelo}
Mastrangelo IA, Bezanilla M, Hansma PK, Hough PV, Hansma HG. 
{\sl Structures of large T antigen at the origin of SV40 DNA replication by atomic force microscopy}. 
Biophys J 1994;66:293-298.

\bibitem{virology}
Roos WH, Bruisma R, Wuite GJL. {\sl Physical virology}. 
Nat Phys 2010;6:733-743. 

\bibitem{virus1}
Cieplak M, Robbins MO.
{\sl Nanoindentation of 35 virus capsids in a molecular model:
Relating Mechanical Properties to Structure}. 
PLOS ONE 2013;8:e63640.

\bibitem{teleschulten}
Lee EH, Gao M, Pinotsis N, Wilmanns M, Schulten K.
{\sl Mechanical strength of the titin Z1Z2-telethonin complex}.
Structure 2006;14:497–509.

\bibitem{telerief}
Bertz M, Wilmanns M, Rief M. 
{\sl The titin-telethonin complex is a directed,
superstable molecular bond in the muscle Z-disk}. 
Proc Natl Acad Sci USA 2009;106:13307-13310.

\bibitem{Dima}
Dima RI, Joshi H. 
{\sl Probing the origin of tubulin rigidity with molecular simulations}. 
Proc Natl Acad Sci USA 2008;105:15743-15748.

\bibitem{cystasikora}
Sikora M, Cieplak M. 
{\sl Mechanical stability of multidomain proteins and novel mechanical clamps}. 
Proteins: Struct Funct Bioinf 2010;79:1786-1799.

\bibitem{cystajaskolski}
Janowski R, Kozak M, Janowska E, Grzonka Z, Grubb A,
Abrahamson M, Jask{\'o}lski M. 
{\sl Human cystatin C, an amyloidogenic protein, dimerizes through
three-dimensional domain swapping}. 
Nat Struct Biol 2001;8:316–320.

\bibitem{Zhmurov}
Zhmurov A, Brown AEX, Litvinov RI, Dima RI, Weisel JW, Barsegov V. 
{\sl Mechanism of Fibrin(ogen) forced unfolding}. 
Structure 2011;19:1615-1624.

\bibitem{Dietz}
Dietz H, Berkemeier F, Bertz M, Rief M.
{\sl Anisotropic deformation response of single protein molecules.}
Proc Natl Acad Sci USA 2006; 103:12724-12728.

\bibitem{dimeric}
Sikora M, Cieplak M.
{\sl Formation of cystine slipknots in dimeric proteins}. 
PLOS ONE 2013;8:e57443.

\bibitem{plugprl}
Sikora M, Cieplak M.
{\sl Cystine plug and other novel mechanisms of large mechanical stability
in dimeric proteins}. 
Phys Rev Lett 2012;109:208101.

\bibitem{liwo}
Cieplak M. 
{\sl Mechanostability of virus capsids and their proteins in structure-based models}. 
In: Liwo A, editor. 
{\it Computational methods to study the structure and dynamics of biomolecules 
and biomolecular processes - from bioinformatics to molecular quantum mechanics}. 
Heidelberg: Springer; 2014. p 295-315.

\bibitem{Velikovsky_2PPT} 
Velikovsky CA, Deng L, Chlewicki LK, Fernandez MM, Kumar V, Mariuzza RA. 
{\sl Structure of Natural Killer Receptor 2B4 Bound to CD48 Reveals Basis
for Heterophilic Recognition in Signaling Lymphocyte Activation Molecule Family}. 
Immunity 2007;27:572–584.

\bibitem{Oberhauser2009} Fuson KL, Ma L, Sutton RB, Oberhauser AF. 
{\sl The C2 Domains of Human Synaptotagmin 1 Have Distinct Mechanical Properties}. 
Biophys J 2009;96:1083-1090.

\bibitem{Takahashi2010} Takahashi H, Shahin V, Henderson RM, Takeyasu K, Edwardson JM.
{\sl Interaction of Synaptotagmin with Lipid Bilayers, Analyzed by Single-Molecule Force Spectroscopy}. 
Biophys J 2010;99:2550-2558. 

\bibitem{Hoang2}
Cieplak M, Hoang TX. 
{\sl Universality classes in folding times of proteins}. 
Biophys J 2003;84:475-488.

\bibitem{JPCM}
Su{\l}kowska JI, Cieplak M. 
{\sl Mechanical stretching of proteins -- a theoretical survey of the Protein Data Bank}. 
J Phys Cond Mat 2007;19:283201. 

\bibitem{models}
Su{\l}kowska JI, Cieplak M. 
{\sl Selection of optimal variants of Go-like models of proteins through studies of stretching}. 
Biophys J 2008;95:3174-3191.

\bibitem{SikoraPLOS2009} 
Sikora M, Su{\l}kowska JI, Cieplak M. 
{\sl Mechanical strength of 17 134 model proteins and cysteine slipknots}. 
PLOS Comp Biol 2009;5:e1000547.

\bibitem{Tsai}
Tsai J, Taylor R, Chothia C, Gerstein M. 
{\sl The packing density in proteins: Standard radii and volumes}. 
J Mol Biol 1999;290:253-266.

\bibitem{Kwiecinska}
Kwiecinska JI, Cieplak M. 
{\sl Chirality and protein folding}. 
J Phys Cond Mat 2005;17:S1565-S1580. 

\bibitem{Thirumalai}
Veitshans T, Klimov D, Thirumalai D.
{\sl Protein folding kinetics:Timescales, pathways and energy landscapes
in terms of sequence dependent properties}. 
Folding and Design 1997;2:1-22. 

\bibitem{flow}
Szymczak P, Cieplak M. 
{\sl Stretching of proteins in a uniform flow}. 
J Chem Phys 2006;125:164903. 

\bibitem{Bell}
Bell G. {\sl Models for the specific adhesion of cells to cells}.
Science 1978;200:618-627.

\bibitem{Evans1997} Evans E, Ritchie K.
{\sl Dynamic Strength of Molecular Adhesion Bonds}.
Biophys J 1997;72:1541-1555.

\bibitem{Dudko2006} Dudko OK, Hummer G, Szabo A. 
{\sl Intrinsic Rates and Activation Free Energies from Single-Molecule Pulling Experiments}. 
Phys Rev Lett 2006;96:108101.

\bibitem{Dima2011} Duan L, Zhmurov A, Barsegov V, Dima RI.
{\sl Exploring the Mechanical Stability of the C2 Domains in Human Synaptotagmin 1}. 
J Phys Chem B 2011;115:10133-10146.

\bibitem{HuWeikl_PNAS_2013} Hu J, Lipowsky R, Weikl TR.
{\sl Binding constants of membrane-anchored receptors and ligands depend 
strongly on the nanoscale roughness of membranes}. 
Proc Natl Acad Sci USA 2013;110:15283-15288.


\end{thebibliography}
\end{document}